\documentstyle[preprint,revtex,eqsecnum]{aps}
\begin{document}
\draft
\baselineskip = 1.5\baselineskip
\begin{title}
Critical behavior of three-dimensional magnets \\ 
with complicated ordering \\
from three-loop renormalization-group expansions
\end{title}
\author{A. I. Sokolov and K. B. Varnashev}
\begin{instit}
Department of Physical Electronics, \\ 
Saint Petersburg Electrotechnical University, \\
Professor Popov Street 5, St.Petersburg, 197376, Russia
\end{instit}

\begin{abstract}
The critical behavior of a model describing phase transitions 
in 3D antiferromagnets with $2N$-component real order 
parameters is studied within the renormalization-group (RG) approach. 
The RG functions are calculated in the three-loop order and resummed 
by the generalized Pad$\acute {{\rm e}}$-Borel procedure preserving
the specific symmetry properties of the model. Anisotropic stable fixed 
point is found to exist in the RG flow diagram for $N \ge 2$ and lie 
near the Bose fixed point; corresponding critical exponents are 
close to those of the $XY$-model. The accuracy of the results obtained 
is discussed and estimated.
\end{abstract}


\narrowtext

\newpage

In this Brief Report, we study within the field-theoretical RG approach in
three dimensions (3D) the critical behavior of a model with
three quartic coupling constants describing certain antiferromagnetic and
structural phase transitions. The Hamiltonian of the model reads 
\begin{equation}
H = \int d^3x\biggl[{\frac 12}(m_0^2\varphi _\alpha \varphi _\alpha +\nabla
\varphi _\alpha \nabla \varphi _\alpha )+{\frac{u_0}{4!}}(\varphi _\alpha
\varphi _\alpha )^2  
+\ {\frac{v_0}{4!}}\varphi _\alpha ^4+{\frac{2z_0}{4!}}\varphi _{2\beta
-1}^2\varphi _{2\beta }^2\biggr], \quad   \label{eq:1.1}
\end{equation}
where ${\bf \varphi _\alpha }$ is a real vector order parameter field, $%
\alpha =1,2,$ \ldots , $2N$, $\beta =1,2,$ \ldots , $N$. For $u_0 = 0$ 
Hamiltonian (\ref{eq:1.1}) describes $N$ non-interacting anisotropic 
$XY$-models, for $z_0 = 0$ it reduces to that of the cubic 
model. When $N=2$ the expression (\ref{eq:1.1}) is relevant to the 
structural phase transition in the $NbO_2$ crystal and, for $v_0 =z_0$, 
to antiferromagnetic transitions in $TbAu_2$ and $DyC_2$. Another physically
interesting case $N=3$ corresponds to antiferromagnetic phase transition in
the $K_2IrCl_6$ crystal and, for $v_0 =z_0$, to those in $TbD_2$ and $Nd$ 
\cite{1}.

This model was studied earlier by the $\epsilon $-expansion technique and 
directly in 3D within the lower (one- and two-loop) perturbative orders 
\cite{1,2}. The main result of those investigations was the existence 
of the nontrivial (anisotropic) stable fixed point for $N\ge 2$ which gives 
rise to a new universality class with a certain set of critical exponents. 
The lower-order approximations, however, are known to 
lead to crude quantitative and, sometimes, to contradictory qualitative 
results, especially for systems with complicated symmetry \cite{3}. 
That is why recently the critical behavior of the model (\ref{eq:1.1}) 
has been analyzed in the third order in $\epsilon $ \cite{4}. 
Investigation of the fixed points stability and the calculation of
marginal dimensionality $N_c$ of the order parameter separating
two different regimes of critical behavior confirmed that the model (\ref
{eq:1.1}) possesses the anisotropic stable fixed point for physically 
interesting cases $N=2$ and $N=3$. However, the eigenvalue exponents 
for this point turned out to be twofold degenerate in the one-loop 
approximation \cite{5}. Such a degeneracy decreases markedly the accuracy 
expected within a given approximation and makes the calculating the 
eigenvalue exponents in higher orders in $\epsilon$ very difficult, while 
resummation of shorter series fails to provide proper numerical estimates. 
It is reasonable therefore to study the model (\ref{eq:1.1}) by means of 
some alternative approach. Below, the critical behavior of this model will 
be investigated by the RG technique in 3D in the three-loop approximation, 
aiming to improve existing qualitative and numerical results.  

The character of the critical behavior is determined by the RG equations 
for quartic coupling constants. Calculating the $\beta$-functions entering 
their right-hand sides and critical exponents $\gamma$ and $\eta$ as 
functions of the dimensionless coupling constants $u$, $v$, and $z$ in the 
three-loop approximation, we obtain: 
\[
\begin{array}{lcl}
\beta _u & = & u \biggl\{1-u-{\frac 1{{N+4}}}\bigl( 3v+z\bigr) 
+{\frac 1{{27(N+4)^2}}}\Bigl[2\bigl( 41\ N+95\bigr) u^2+300uv+100uz \\ 
& + & 69v^2+23z^2\Bigr] -{\frac 1{{8(N+4)^3}}}\Bigl[ 
\bigl(5.39577\ N^2+109.88075 N+199.64042\bigr) u^3 \\
& + & \bigl( 39.88127\ N+493.84155\bigr) u^2v+\bigl( 13.29376\ N  
+164.61385\bigr) u^2z \\
& + & \bigl( 3.73134\ N+302.86778\bigr) uv^2
+58.35119uvz +\bigl( 1.24378\ N +81.50553\bigr) \\ 
& \times & uz^2 + 65.93728v^3+3.73134v^2z
+21.97909vz^2+9.35389z^3\Bigr] \biggr\},
\end{array}
\]
\begin{equation}
\begin{array}{lcl}
\beta _v & = & v-{\frac 1{{N+4}}}\bigl( 6uv+{\frac 92}v^2+{\frac 12}z^2 \bigr)
+{\frac 1{{27(N+4)^2}}}\Bigl[ 2\bigl( 23\ N+185\bigr) u^2v + 624uv^2 \\ 
& + & 46uvz+54uz^2+231v^3+23vz^2+18z^3\Bigr] -{\frac 1{{8(N+4)^3}}} 
\Bigl[\bigl( -5.00443\ N^2 \\
& + & 83.70780\ N+469.33398\bigr) u^3v +\bigl( 4.47812\ N 
+1228.60593\bigr) u^2v^2 \\
& + & \bigl( -7.50664\ N+115.06965\bigr) u^2vz+%
\bigl( 2.99978\ N +98.15522\bigr) u^2z^2 \\
& + & 957.78168uv^3+4.47812uv^2z+135.72564uvz^2 +60.68074uz^3 \\
& + & 255.92974v^4+4.10473v^2z^2+49.83176vz^3 +5.05072z^4\Bigr],
\label{eq:2.3}
\end{array}
\end{equation}
\[
\begin{array}{lcl}
\beta _z & = & z \biggl\{1-{\frac 1{{N+4}}}\bigl( 6u+3v+2z\bigr) 
+{\frac 1{{27(N+4)^2}}}\Bigl[2\bigl( 23\ N+185\bigr) u^2 +462uv \\ 
& + & 262uz+69v^2+162vz+41z^2\Bigr] -{\frac 1{{8(N+4)^3}}} \Bigl[\bigl( 
-5.00443\ N^2+83.70780\ N \\
& + & 469.33398\bigr) u^3+\bigl( -4.52123\ N + 934.14027\bigr) u^2v
+\bigl(4.49249\ N \\
& + & 507.69053\bigr) u^2z+411.65505uv^2 +550.60474uvz+196.40638uz^2 \\
& + & 65.93728v^3+108.78727v^2z +118.33643vz^2+21.85596z^3\Bigr] \biggr\};
\end{array}
\]

\begin{eqnarray}
\gamma ^{-1} &=&1-{\frac 1{{4(N+4)}}}\Bigl[2(N+1)u+3v+z\Bigr]+{\frac 1{{%
4(N+4)^2}}}\Bigl[2(N+1)u^2 +6uv+2uz+3v^2+z^2\Bigr] \nonumber \\
& - &{\frac{0.04813}{{(N+4)^3}}}\Bigl%
[4(N+1)^2u^3+18(N+1)u^2v+6(N+1)u^2z+6(N+4)uv^2+12uvz \nonumber \\
& + & 2(N+2)uz^2+9v^3+3v^2z+3vz^2+z^3\Bigr]-{\frac{0.06182}{{(N+4)^3}}}\Bigl%
[4(N+1)(N+4)u^3  \nonumber \\
& + & 18(N+4)u^2v +6(N+4)u^2z+81uv^2+18uvz+21uz^2+27v^3+9vz^2+4z^3\Bigr],  
\label{eq:2.4}
\end{eqnarray}
\begin{eqnarray}
\eta & = &{\frac 2{{27(N+4)^2}}}\Bigl[2\bigl( N+1\bigr) u^2+6uv+2uz+3v^2
+z^2 \Bigr]+{\frac{0.00309}{{(N+4)^3}}} \Bigl[4(N+1)(N+4)u^3  \nonumber \\
& + & 18(N+4)u^2v+6(N+4)u^2z+81uv^2 
+18uvz+21uz^2+27v^3+9vz^2+4z^3\Bigr].  
\label{eq:2.5}
\end{eqnarray}

To extract the physical information from these divergent series, the 
Pad$\acute {{\rm e}}$-Borel resummation procedure will be used. Since the 
expansions of quantities depending on three variables $u$, $v$, and $z$ 
are dealt with, the Borel transformation is taken in a generalized form: 
\begin{equation}
f(u,v,z)=\sum_{ijk}c_{ijk}u^iv^jz^k=\int\limits_0^\infty
e^{-t}F(ut,vt,zt)dt, \qquad
F(x,y,w)=\sum_{ijk}{\frac{c_{ijk}x^i y^j w^k}{{(i+j+k)!}}}. 
\label{eq:3.2}
\end{equation}
To perform an analytical continuation, we address to the resolvent series
\begin{equation}
\tilde F(x,y,w,\lambda )=\sum_{n=0}^\infty \lambda
^n\sum_{l=0}^n\sum_{m=0}^{n-l}{\frac{c_{l,m,n-l-m}x^ly^mw^{n-l-m}}{{n!}}}
\label{eq:3.3}
\end{equation}
which is a series in powers of $\lambda $ with coefficients being uniform
polynomials in $u$, $v$, $z$, and then use Pad$\acute {{\rm e}}$
approximants $[L/M]$ in $\lambda $ at $\lambda = 1$. The 
approximant [3/1] is employed for resummation of 
three-loop RG series Eqs.~(\ref{eq:2.3}). The coordinates of all fixed 
points are determined for two most interesting cases $N=2$ and $N=3$. 
The results obtained are presented in Table I, which contains also, 
for comparison, analogous estimates found earlier from the two-loop RG 
series \cite{2}. The global structure of the RG 
flows in the three-loop approximation is shown in Fig.1 (see the printed 
paper: Phys. Rev. B 59 (1999) 8363-8366).

The three-loop contributions to the $\beta$-functions are seen to affect 
the location of the fixed points considerably. Of special interest is the
fortune of the nontrivial, anisotropic fixed point 8. In the one-loop 
approximation for $N=2$ it coincides with the Heisenberg fixed point 2 
\cite{6}. In higher orders this degeneracy is lifted out and the point 8, 
remaining a stable one, moves toward the Bose fixed point 5 lying within 
the plane $v=z$. With increasing $N$, the former becomes closer and 
closer to the latter, indicating the tendency of the $O(2N)$-symmetric 
interaction to vanish at criticality. This resembles the behavior of the 
cubic model that is known to split into $n$ non-interacting Ising models 
when $n\to \infty $.

The resummation procedure changes the perturbative expansions 
for $\beta$-functions by the complicated non-polynomial expressions. 
Does this procedure preserve the symmetry properties of the system? 
The model (\ref{eq:1.1}) possesses, apart from usual, the special symmetry 
properties which have been already used for testing a validity of 
approximations employed \cite{2}. For example, if the field 
$\varphi _\alpha $ undergoes the transformation 
$\varphi _{2\beta-1} \rightarrow {\frac 1{{\sqrt{2}}}}~(\varphi
_{2\beta-1}+\varphi _{2\beta}),~  
\varphi _{2\beta} \rightarrow {\frac 1{{\sqrt{2}}}}~(\varphi
_{2\beta-1}-\varphi _{2\beta})$,
the coupling constants are also transformed: 
\begin{equation}
u\rightarrow u,\quad v\rightarrow {\frac 12}(v+z),\quad z\rightarrow {\frac
12}(3v-z)\ \ ,  \label{eq:3.14}
\end{equation}
but the structure of the Hamiltonian itself remains the same. Since such 
a transformation does not affect the RG equations, it can, at most, 
rearrange the fixed points leaving their coordinates unchanged \cite{2,3,4}. 
To check up whether such an invariance holds for the fixed points
found, let us apply the transformation (\ref{eq:3.14}) to the content
of Table I. Then points 1, 2, 5 and 8 will stay at their places while the
coordinates of points 3 and 4 will turn into those of points 6 and 7
respectively and vice versa with accuracy of order of $10^{-4}$. Since the 
fixed point coordinates were evaluated numerically with just the same 
accuracy, it means that our resummation procedure exactly reproduces the 
symmetry properties discussed.

Let us determine further the critical exponents. The coefficients of 
the series for $\eta $ rapidly diminish, therefore its value may be 
found by direct substitution of the fixed point coordinates into the 
expansion (\ref{eq:2.5}). In the case of
susceptibility exponent corresponding series (\ref{eq:2.4})
have to be resummed. Using the generalized Pad$\acute {{\rm e}}$-Borel
procedure, we evaluate the exponent $\gamma $ and then,
addressing the values of $\eta $, calculate critical exponent $\alpha $ 
by the scaling relations. The results for $N=2$ and $N=3$
are presented in Table II. As is seen, the critical exponents of the fixed 
point 8 only slightly differ from those of the Bose one reflecting a 
closeness of both fixed points in the $(u, v, z)$ space.

The critical exponents for the fixed points 2 (Heisenberg), 5 (Bose), 
and 3 (Ising) thus found are close to their high-precision 
analogs resulting from the six-loop RG expansions \cite{8}; the 
differences do not exceed 0.02-0.03. 
For the anisotropic fixed point 8 the situation is more complicated.
In the three-loop approximation it is three-dimensionally stable and
characterized by negative value of the specific heat exponent $\alpha$,
while the Bose fixed point looks
unstable since it has a positive $\alpha $. In fact, however, the 
$XY$-like critical behavior in 3D is described by the negative 
$\alpha $ as is known both from the six-loop RG calculations \cite{8} and
recent extremely accurate measurements \cite{9}. Moreover, a sign of the
exponent $\alpha $ at the Bose fixed point determines either the 
$O(2N)$-symmetric interaction is relevant near this point or not \cite{10}: 
if $\alpha <0$, the Bose fixed point should be stable with respect to this
interaction. It means that what is really stable is
the fixed point 5 while the stability of the fixed point 8 within the plane 
$v=z$ is an artifact of the three-loop approximation.

Keeping this in mind we can estimate an accuracy of the
anisotropic fixed point coordinates given by the three-loop RG series. 
Because of the obvious topological reasons (see Fig.1), the
point 8, being in fact unstable, should have $u_c < 0$. At the same time,
since the true value of $\alpha $ at the Bose fixed point is very small 
\cite{8,9} the points 5 and 8 should be very close one to another
and the modulus of $u_c$ for the point 8 should be very small as well. 
Hence, as can be deduced from the numbers in the last column 
of Table I, the three-loop approximation predicts locations of 
this point for $N=2$ and $N=3$ with errors about 0.2 and 0.1 respectively.

Dealing with the theory without a small parameter and the short
(three-loop) perturbative series one would refer to such an accuracy
as satisfactory. On the other hand, numerically small errors lead in this
case to qualitatively incorrect results making the situation rather 
unfavorable. The point is that in 3D the model (\ref{eq:1.1}) is
almost identical to some marginal system for which $\alpha =0$ at the 
$XY$-like criticality and, as a consequence, fixed points 5 and 8 coincide. 
Since such a marginality is far from to be seen in the one-loop 
approximation, it hardly manifests itself within a perturbation theory,
even in higher orders.

The "near-marginality" is not a unique feature of the model (\ref{eq:1.1}) 
being typical for 3D systems with several coupling constants. Thus, for
the cubic model the marginal value of $n$ that separates
two different regimes of critical behavior is numerically close
to the physical value $n=3$ \cite{11}. The model describing 
phase transitions into chiral states has the marginal dimensionality of the
complex order parameter $N_{c2}$ (separating domains of continuous and
first-order transitions) that is also very close to the physical value 
$N=2$ \cite{3}, etc. Hence, to answer the question about the type of the 
critical behavior of such ''unconvenient'' models the
higher-order RG analysis should be carried out.

The calculation of the RG functions of the model (\ref{eq:1.1}) in the next, 
four-loop approximation is an extremely difficult problem. The point
discussed, however, may be cleared up without performing additional RG 
calculations. Indeed, let us trace how the numerical value of the exponent 
$\alpha $ for the Bose fixed point depends on the order of the RG 
approximation. The estimates for $\alpha $ obtained in the one-,
two-, three-, four-, five-, and six-loop orders using the 
Pad$\acute {{\rm e}}$-Borel resummation technique are 0.125, --0.012, 
0.009, --0.007, --0.007, and --0.007, respectively. 
As is seen, the sign of exponent $\alpha $ alternates up to four-loop order 
and then stays negative indicating the stability of the Bose fixed point. 
Since two neighbouring fixed points can not be stable 
simultaneously we conclude that the fixed point 8 should be unstable within 
the four-loop and higher-order approximations. Consequently, the next, 
four-loop RG approximation will be sufficient to yield the correct structure 
of the RG flow diagram as well as high-precision numerical estimates for 
the critical exponents.

\newpage

\figure{Three-dimensional flow diagram of the RG equations for $N = 2$ in
the three-loop approximation.} \label{Fig.1}

\newpage

\widetext
\begin{table}
\caption{Coordinates of the fixed points of the RG equations for $N = 2$ 
and $N = 3$ obtained within three-loop ([3/1]) and two-loop ([2/1]) 
approximations.}
\label{table1}
\begin{tabular}{cccccccccc}
&  & 1 & 2 & 3 & 4 & 5 & 6 & 7 & 8 \\ 
\tableline
N = 2 \\
\tableline 
$u_c$ & 
$[3/1]$ & 0.0 & 1.3671 & 0.0 & 0.9254 & 0.0 & 0.0 & 0.9255 & 0.1872 \\ & 
$[2/1]^{*}$ & 0.0 & 1.4863 & 0.0 & 1.0166 & 0.0 & 0.0 & 1.0166 & 0.0332 \\ 
\tableline 
$v_c$ & 
$[3/1]~$ & 0.0 & 0.0 & 1.8883 & 0.7764 & 1.6833 & 0.9442 & 0.3882 & 1.4914 \\ 
& $[2/1]^{*}$ & 0.0 & 0.0 & 2.1289 & 0.8358 & 1.8700 & 1.0645 & 0.4178 & 
1.8344 \\ 
\tableline 
$z_c$ & 
$[3/1]~$ & 0.0 & 0.0 & 0.0 & 0.0 & 1.6833 & 2.8325 & 1.1646 & 1.4914 \\ & 
$[2/1]^{*}$ & 0.0 & 0.0 & 0.0 & 0.0 & 1.8700 & 3.1934 & 1.2537 & 1.8344 \\ 
\tableline
N = 3 \\
\tableline 
$u_c$ & 
$[3/1]~$ & 0.0 & 1.3310 & 0.0 & 0.6005 & 0.0 & 0.0 & 0.6005 & 0.0780 \\ & 
$[2/1]^{*}$ & 0.0 & 1.4262 & 0.0 & 0.6307 & 0.0 & 0.0 & 0.6307 & 0.0094 \\ 
\tableline 
$v_c$ & 
$[3/1]~$ & 0.0 & 0.0 & 2.2030 & 1.4971 & 1.9639 & 1.1015 & 0.7485 & 1.8845 \\ 
& $[2/1]^{*}$ & 0.0 & 0.0 & 2.4837 & 1.6855 & 2.1816 & 1.2419 & 0.8428 & 
2.1716 \\ 
\tableline 
$z_c$ & 
$[3/1]~$ & 0.0 & 0.0 & 0.0 & 0.0 & 1.9639 & 3.3045 & 2.2456 & 1.8845 \\ & 
$[2/1]^{*}$ & 0.0 & 0.0 & 0.0 & 0.0 & 2.1816 & 3.7256 & 2.5283 & 2.1716 \\ 
\end{tabular}
\tablenotes{$^{*}$ Quoted from Ref.~\cite{2}}
\end{table}

\widetext
\begin{table}
\caption{Critical exponents for $N = 2$ (upper lines) and $N=3$ (lower
lines) calculated within three-loop approximation.}
\label{table3}
\begin{tabular}{ccccccccc}
& 1 & 2 & 3 & 4 & 5 & 6 & 7 & 8 \\ 
\tableline $\gamma$ & 1 & 1.4260 & 1.2406 & 1.3989 & 1.3098 & 1.2406 & 1.3990
& 1.3360 \\ 
& 1 & 1.5164 & 1.2406 & 1.4080 & 1.3098 & 1.2406 & 1.4080 & 1.3291 \\ 
\tableline $\eta$ & 0.0 & 0.0257 & 0.0246 & 0.0258 & 0.0260 & 0.0246 & 0.0258
& 0.0261 \\ 
& 0.0 & 0.0238 & 0.0246 & 0.0253 & 0.0260 & 0.0246 & 0.0253 & 0.0261 \\ 
\tableline $\alpha$ & 0.0 & -0.1669 & 0.1159 & -0.1258 & 0.0094 & 0.1159 & 
-0.1259 & -0.0305 \\ 
& 0.0 & -0.3020 & 0.1159 & -0.1390 & 0.0094 & 0.1159 & -0.1390 & -0.0199 \\ 
\end{tabular}
\end{table}


\begin{references}
\bibitem{6}  In the one-loop approximation for $N=2$ the Heisenberg fixed
point is fourfold degenerate \cite{2}. So strong degeneracy, being typical 
for a number of complicated models \cite{3,7}, is in fact 
occasional and lifted out in the higher-order approximations.
\end{references}
\end{document}